\documentclass[preprint,showpacs,preprintnumbers,amsmath,amssymb]{revtex4}

\usepackage{graphicx}
\usepackage{dcolumn}
\usepackage{amsmath}
\usepackage{amssymb}

\newcommand{\bvec}[1]{\mbox{\boldmath $#1$}}
\newcommand{\D}{\delta }

\def\vq{{\bf q}}

\def\vr{{\bf r}}
\def\vS{{\bf S}}

\newcommand{\fig}[1]{Fig.~\ref{#1}}
\newcommand{\be}{\begin{equation}}
\newcommand{\ee}{\end{equation}}
\newcommand{\bea}{\begin{eqnarray}}

\newcommand{\eea}{\end{eqnarray}}
\newcommand{\bean}{\begin{eqnarray*}}
\newcommand{\eean}{\end{eqnarray*}}
\newcommand{\bfi}{\begin{figure}}
\newcommand{\efi}{\end{figure}}
\newcommand{\bc}{\begin{center}}
\newcommand{\ec}{\end{center}}
\newcommand{\ba}{\begin{array}}
\newcommand{\ea}{\end{array}}

\begin{document}


\title{Cuprate Superconductors in the Vicinity of a Pomeranchuk Instability}

\author{Hiroyuki Yamase} 
\affiliation{Max-Planck-Institute for Solid State Research, 
Heisenbergstrasse 1, D-70569 Stuttgart, Germany}





\begin{abstract} 
We propose that cuprate superconductors are in the vicinity of 
a spontaneous $d$-wave type Fermi surface symmetry breaking, 
often called a $d$-wave Pomeranchuk instability. 
This idea is explored 
by means of a comprehensive study of 
magnetic excitations within the slave-boson 
mean-field theory of the $t$-$J$ model. 
We can naturally understand the pronounced 
$xy$ anisotropy of magnetic excitations in untwinned 
YBa$_{2}$Cu$_{3}$O$_{y}$ 
and the sizable change of incommensurability 
of magnetic excitations at the transition 
temperature to the low-temperature tetragonal lattice structure 
in La$_{2-x}$Ba$_{x}$CuO$_{4}$. 
In addition, the present theoretical framework allows the understanding  
of the similarities and differences of magnetic excitations in 
Y-based and La-based cuprates. 
\end{abstract}

\maketitle

\section{Introduction} 
Usually the symmetry of the Fermi surface (FS) satisfies 
the point-group symmetry of the underlying lattice structure. 
However, recently, symmetry breaking of the FS with the 
$d$-wave order parameter was discussed in the $t$-$J$\cite{yamase00} and 
Hubbard\cite{metzner00} models on a square lattice: 
the FS expands along the $k_{x}$ direction 
and shrinks along the $k_{y}$ direction, or vice versa. 
This $d$-wave type Fermi surface deformation  ($d$FSD) is often 
called a $d$-wave Pomeranchuk instability, referring to 
Pomeranchuk's stability criteria of isotropic 
Fermi liquids.\cite{pomeranchuk58} 
But the $d$FSD occurs even without violating Pomeranchuk's criteria 
since the transition is usually of first order at 
low temperature,\cite{khavkine04} and also occurs 
in strongly correlated electron  systems such as 
those described by the $t$-$J$ model. 

The $d$FSD competes with superconductivity. 
In the slave-boson mean-field theory of the $t$-$J$ model\cite{yamase00} 
superconductivity becomes more dominant and the $d$FSD instability 
does not occur, which was recently confirmed by  
a variational Monte Carlo calculation.\cite{edegger06} 
However, substantial correlations of the $d$FSD still 
remain and lead to a giant response to a 
small anisotropy, e.g. in a lattice.\cite{yamase00,edegger06} 

Recently a pronounced anisotropy was reported in 
magnetic excitation spectra of untwinned 
YBa$_{2}$Cu$_{3}$O$_{y}$ (YBCO).\cite{hinkov04,hinkov07} 
In the superconducting state\cite{hinkov04} the peak of the 
imaginary part of the dynamical magnetic susceptibility $\chi(\vq,\,\omega)$ 
at various frequencies appear at 
$\vq =(\pi\pm 2\pi\eta_{x},\pi)$ and $(\pi,\pi\pm 2\pi\eta_{y})$ 
with $\eta_{x}\neq \eta_{y}$ and different peak intensity; 
$\eta_{x(y)}$ parameterizes the degree of incommensurability. 
In the pseudogap phase\cite{hinkov07} the incommensurate (IC) peaks are 
smeared, and the spectral weight forms a broad distribution 
centered at 
$(\pi,\pi)$, but with a strongly enhanced anisotropy characterized by 
an elliptic shape. 
In La-based cuprates, IC signals appear up to a temperature much higher 
than the superconducting transition temperature $T_{c}$. 
While the IC signals retain fourfold symmetry, 
namely $\eta_{x}=\eta_{y}=\eta$, 
the value of $\eta$ turns out to show a sizable change 
when the lattice undergoes the structural phase transition from 
the low-temperature orthorhombic structure (LTO) to the 
low-temperature tetragonal structure (LTT) in 
La$_{2-x}$Ba$_{x}$CuO$_{4}$ (LBCO) with $x=0.125$.\cite{fujita04}

In this paper, we show that the above experimental data 
are naturally understood in terms of $d$FSD correlations. 
While sizable $d$FSD correlations were found in 
the slave-boson,\cite{yamase00} 
exact diagonalization,\cite{miyanaga06} and 
variational Monte Carlo\cite{edegger06} 
techniques in the $t$-$J$ model, and 
in various renormalization 
group analyses\cite{metzner00,wegner02,honerkamp02,kampf03} 
in the Hubbard model, 
we employ the $t$-$J$ model in the slave-boson mean-field theory, 
which has the advantage of dealing with both $d$-wave singlet pairing 
and $d$FSD correlations on an equal footing as well as 
performing a systematic calculation of magnetic excitations.

\section{Model and formalism}
The $t$-$J$ model 
\be
 H = -  \sum_{\vr,\,\vr',\, \sigma} t^{(l)}_{\boldsymbol \tau}  
 \tilde{c}_{\vr\,\sigma}^{\dagger}\tilde{c}_{\vr'\,\sigma}+
    \sum_{\langle \vr,\vr' \rangle} \, J_{\boldsymbol \tau} 
 \vS_{\vr} \cdot \vS_{\vr'}  \label{tJ} 
\ee  
is defined in the Fock space with no doubly occupied sites. 
The operator $\tilde{c}_{\vr\,\sigma}^{\dagger}$
($\tilde{c}_{\vr\,\sigma}$) creates (annihilates) an electron with
spin $\sigma$ on site $\vr$, and $\vS_{\vr}$ is the 
spin operator; 
$J_{\boldsymbol \tau}(>0)$ is a superexchange coupling between the 
nearest neighbor sites and $t^{(l)}_{\boldsymbol \tau}$  is a 
hopping amplitude between the $l$th nearest neighbors $(l\leq 3)$; 
${\boldsymbol \tau}=\vr'-\vr$.

We introduce the slave particles, $f_{\vr \sigma}$ and $b_{\vr}$, 
as $\tilde{c}_{\vr \sigma}=b_{\vr}^{\dagger}f_{\vr \sigma}$,  
where $f_{\vr \sigma}$ ($b_{\vr}$) is a fermion (boson) operator 
that carries spin $\sigma$ (charge $e$), and $\vS_{\vr}=\frac{1}{2}
f_{\vr \alpha}^{\dagger}{\bvec \sigma}_{\alpha \beta} f_{\vr \beta}$ 
with the Pauli matrices ${\bvec \sigma} = (\sigma^x,\sigma^y,\sigma^z)$.
The slave bosons and fermions are linked by the local constraint
$b_{\vr}^{\dagger} b_{\vr} + 
 \sum_{\sigma} f_{\vr \sigma}^{\dagger} f_{\vr \sigma} = 1$.
This is an exact transformation known as the slave-boson formalism. 
We then decouple the interaction with the so-called 
resonating-valence-bond  (RVB) mean fields: 
$\chi_{\boldsymbol \tau}$$\equiv$$\langle \sum_{\sigma}f_{\vr\,\sigma}^{\dagger}
f_{\vr' \,\sigma}\rangle$, 
$\langle b_{\vr}^{\dagger}b_{\vr'}\rangle$, and 
$\Delta_{\boldsymbol \tau}$$\equiv$$\langle f_{\vr\,\uparrow}f_{\vr' \,\downarrow}- 
f_{\vr\,\downarrow}f_{\vr' \,\uparrow}\rangle$. 
These mean fields are assumed to be real constants 
independent of site $\vr$. 
We approximate the bosons to condense at the bottom of the band, which 
leads to 
$\langle b_{\vr}^{\dagger}b_{\vr'}\rangle=\D$, where $\D$ is the hole 
density. 
The order parameter of the $d$FSD is defined as 
$\phi=(\chi_{y}-\chi_{x})/2$ and  
$d$-wave singlet pairing is given by 
$\Delta=|\Delta_{x}-\Delta_{y}|/2$; 
both orders are generated by the second term in the Hamiltonian (1). 
After determining the RVB mean fields 
by minimizing the free energy, we compute the dynamical magnetic 
susceptibility $\chi(\vq,\,\omega)$  numerically 
in the renormalized random phase 
approximation (RPA).\cite{brinckmann99,yamase99}

The material dependence of high-$T_{c}$ cuprates is taken into account 
mainly by different choices of band parameters. 
While the parameters $t^{(l)}_{\boldsymbol \tau}$ and 
$J_{\boldsymbol \tau}$ are considered on a square lattice for La-based 
cuprates, the bilayer coupling is included to perform a more 
realistic calculation for Y-based cuprates. 
The details of our formalism are presented 
in Refs.~\onlinecite{yamase07c} and \onlinecite{yamase07} 
for Y-based and La-based cuprates, respectively. 
We show results for $\delta=0.12$, which  may be 
complement to those references.

\section{Results for untwinned Y-based cuprates}
The crystal structure of untwinned YBCO is orthorhombic in the 
carrier doping region where superconductivity is realized at low $T$, 
yielding a small $xy$ anisotropy to the electronic band structure. 
We introduce $5\%$ anisotropy to hopping integrals between the $x$ and $y$ 
direction, and $10\%$ anisotropy between $J_{x}$ and $J_{y}$, 
twice as large, as imposed by the superexchange mechanism. 
Since CuO chains run along the $y$ direction, we assume the band 
parameters are enhanced more along the $y$ direction.

The order parameter of the $d$FSD $\phi$ is plotted 
in \fig{mean-field}(a) together with the $d$-wave singlet pairing  
$\Delta$. 
$\phi$ increases with decreasing $T$ 
and exhibits a cusp at the onset of 
$\Delta$, which is denoted as $T_{\rm RVB}$; $T_{\rm RVB}$ is 
interpreted as pseudogap crossover temperature $T^{*}$ in the underdoped 
regime and as $T_{c}$ in the overdoped regime of high-$T_{c}$ cuprates. 
Below $T_{\rm RVB}$, $\phi$ is suppressed because of 
competition with singlet pairing but still enhanced 
compared with the value of $\phi$ at high $T$. 

The FSs at low $T$ are shown in \fig{mean-field}(b). The inner (outer) FS 
is formed by the anti-bonding (bonding) band due to the bilayer coupling. 
The inner FS can easily open in the presence of an anisotropy 
while the outer FS is still holelike. 

Momentum space maps of Im$\chi(\vq,\,\omega)$ for a sequence of 
temperatures are shown in \fig{q-maps} for 
$\omega=0.30J$ and $\delta=0.12$ in the odd channel; 
a similar result is obtained 
in the even channel.\cite{yamase07c} 
The strong intensity region forms a deformed diamond shape at 
low $T$ [\fig{q-maps}(a)], and the spectral weight inside the diamond 
gradually increases with $T$ [\fig{q-maps}(b)]. 
The spectral weight around $\vq=(\pi,\,\pi)$ becomes dominant 
at temperature slightly below $T_{\rm RVB}=0.124J$ and 
an enhanced anisotropy appears with an elliptic-shape 
distribution  as shown in \fig{q-maps}(c). 
When $T$ increases further, however, the anisotropy is  
reduced [\fig{q-maps}(d)]. This strong $T$ dependence 
is characteristic of the effect of $d$FSD correlations.

\section{Results for La-based cuprates}
Three different crystal structures are realized in La-based high-$T_{c}$ 
cuprates, depending on temperature and carrier doping: 
high-temperature tetragonal structure (HTT), LTO, and LTT. 
In the former two structures, the lattice does not produce 
a $xy$ anisotropy and thus not couple to the underlying $d$FSD tendency. 
We expect an electronlike FS for $\delta=0.12$ as shown in \fig{FS}(a). 
The LTT, however, yields a small $xy$ anisotropy, 
the direction of which alternates along the $z$ axis. 
Such a small anisotropy is then strongly enhanced by $d$FSD correlations 
as we have seen in \fig{mean-field}(a).  
Through a coupling 
to the LTT, therefore, we expect a strongly deformed FS as shown 
by solid lines in \fig{FS}(b); the FS is deformed in the opposite 
direction (gray lines) in neighboring CuO$_{2}$ planes and thus 
the superimposed FS recovers fourfold symmetry. 
A weak interlayer coupling then leads to 
two FSs, an inner electronlike FS and 
an outer holelike FS as shown in \fig{FS}(c). 
Note that dynamical fluctuations of the $d$FSD are expected 
in the presence of the soft phonon mode toward the LTT phase transition 
even in the LTO phase.\cite{thurston89b} 

For simplicity we do not consider the interlayer coupling, although 
it is important in the sense that 
it can yield a one-dimensional-like incommensurate 
peak along the direction $\vq=\frac{1}{\sqrt 2}(q,q)$.\cite{yamase02} 
We compute Im$\chi(\vq,\,\omega)$ for the FSs shown in 
Figs.~\ref{FS}(a) and (b). We find that 
similar magnetic excitations are obtained for both FSs.\cite{yamase07}   
The spectral weight distribution has fourfold symmetry 
around $\vq=(\pi,\,\pi)$. 
At low $T$, the strong intensity region forms a diamond shape as shown in 
\fig{q-maps}(a), and pronounced IC signals appear at 
$(\pi\pm 2\pi\eta,\,\pi)$ and $(\pi,\,\pi\pm 2\pi\eta)$. 
With increasing $T$, the spectral weight around $\vq=(\pi,\,\pi)$ 
increases and the IC signals become less clear, but they 
are still discernible even at high 
temperature ($\sim 0.15J$),\cite{yamase07} 
different from the result in \fig{q-maps}(d). 
Since $d$FSD correlations 
produce a distinct change of the FS shape from \fig{FS}(a) to (b)  
when the lattice undergoes a phase transition to the LTT, 
such a FS change in general leads to a sizable change of 
$\eta$ at the LTT transition. This features the underlying $d$FSD 
correlations in La-based cuprates.

\section{Discussion and conclusion} 
\begin{figure}
\centerline{\includegraphics[width=0.45\textwidth]{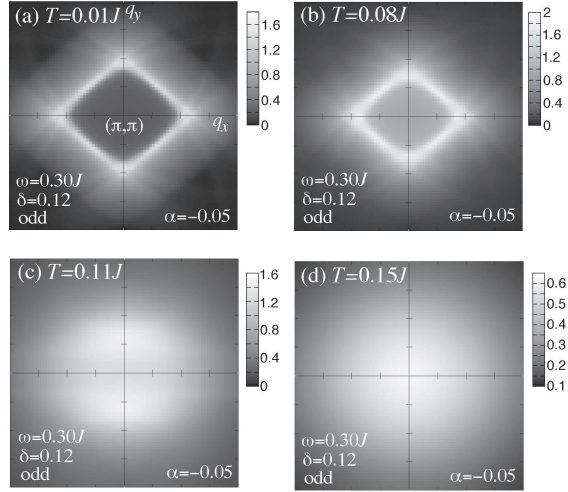}}
\caption{$\vq$ maps of Im$\chi(\vq,\omega)$ for a sequence of $T$ 
in $0.6\pi \leq q_{x},q_{y} \leq 1.4\pi$ for 
$\omega=0.30J$ and $\delta=0.12$ in the odd channel; here $T_{\rm RVB} = 0.124J$. 
} 
\label{q-maps}
\end{figure}

\begin{figure}
\centerline{\includegraphics[width=0.4\textwidth]{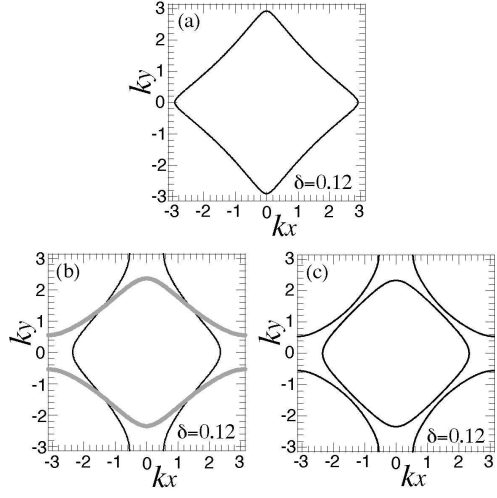}}
\caption{(a) Fermi surface expected in the HTT and LTO phases  
for $\delta=0.12$. 
(b) and (c) Fermi surface expected in the LTT phase; 
a weak interlayer coupling is considered in (c) but not in (b). 
}
\label{FS}
\end{figure}

$d$FSD correlations lead to a strong anisotropy of 
magnetic excitations at relatively high $T$ [\fig{q-maps}(c)], 
which accounts for the recent observation in the pseudogap phase in 
untwinned YB$_{2}$Cu$_{3}$O$_{6.6}$.\cite{hinkov07} 
In La-based cuprates, $d$FSD correlations are expected to lead to a 
change of the FS (\fig{FS}) through the LTT phase transition, 
which is in general accompanied by a sizable change of incommensurability 
of magnetic excitations; the FS shown in \fig{FS}(b) tends to favor 
a larger $\eta$. 
Such a change was recently observed in LBCO 
with $x=0.125$,\cite{fujita04} although the authors\cite{fujita04} 
interpreted the data differently as due to a charge stripe formation.

Magnetic excitations in high-$T_{c}$ cuprates, especially in La-based 
cuprates, are frequently discussed in terms of charge stripes 
or electronic nematic order envisaged as a partial 
stripe order.\cite{kivelson03} 
Although the state with a $d$-wave deformed FS has the same symmetry 
as the nematic order, $d$FSD correlations originate from forward 
scattering processes of quasi-particles\cite{yamase00,metzner00} 
and do not require charge stripe correlations.  
An interesting open question is whether the $d$FSD state can 
have an instability 
toward a charge ordered state such as stripes. 
Even if it were the case, our calculation shows that 
many salient features of magnetic excitations 
observed in La-based cuprates 
are already well-captured without stripes. 
Observations of weak charge order signals\cite{tranquada95} 
do not necessarily mean that charge stripes are crucial to magnetic 
excitations. 
The effect of charge order on magnetic excitations can be higher 
order corrections beyond the present renormalized RPA.

\begin{figure}
\centerline{\includegraphics[width=0.45\textwidth]{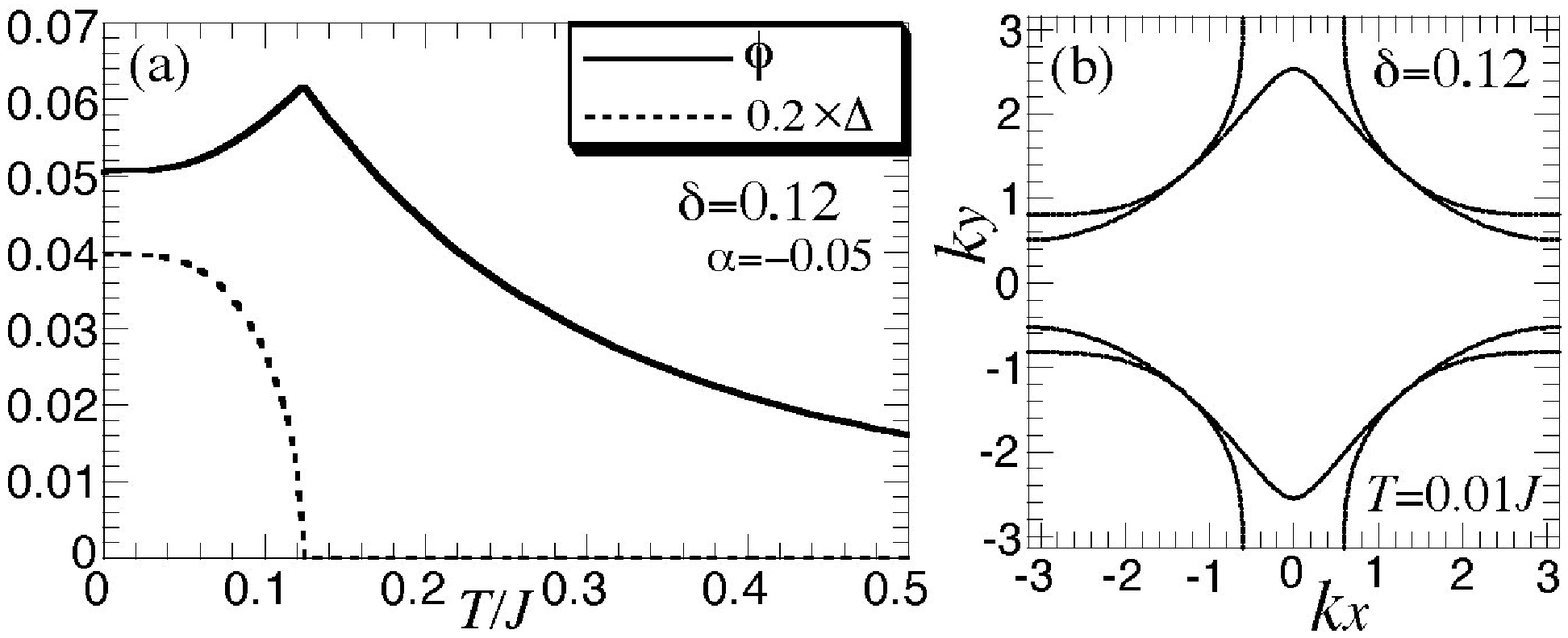}}
\caption{The mean-field solution in the presence of an anisotropy 
for $\delta=0.12$. (a) $T$ dependence of $\phi$ and $\Delta$. 
(b) Fermi surface at low $T$. 
}
\label{mean-field}
\end{figure}

A well-known distinction of magnetic excitations 
between La-based and Y-based cuprates is that 
low-energy IC signals are realized 
up to a temperature much higher than $T_c$ in the former,\cite{thurston89,aeppli97} while they are realized only below $T_{c}$ or 
possibly below pseudogap temperature $T^{*}$ in the latter.\cite{dai98}   
This difference comes from a FS difference in the present theory. 
For the FSs shown in Figs.~\ref{FS}(a) and (b), there are no 
particle-hole scattering processes with $\vq=(\pi,\,\pi)$ for low energy, 
yielding a robust incommensurate structure even in the normal state, 
while the FS shown in \fig{mean-field}(b) 
allows such scattering processes, 
smearing incommensurate signals in the normal state. 
The FS geometry is essential for the material dependence of 
magnetic excitations, the same insight as 
the early proposal,\cite{si93,tanamoto93} 
although we have extended the early studies in the 
$t$-$J$ model\cite{tanamoto94,li02,brinckmann02} 
by introducing an idea of the $d$FSD. 
More detailed comparisons of magnetic excitations  between Y-based and 
La-based cuprates are presented in Ref.~\onlinecite{yamase07}. 

We have shown that many salient features of magnetic excitations in 
high-$T_{c}$ cuprates are well-captured in terms of particle-hole 
excitations. In particular, $d$FSD correlations are essential for 
the pronounced anisotropy observed in 
untwinned YBCO\cite{hinkov04,hinkov07} and the sizable change of 
incommensurability at the onset temperature of the LTT 
phase transition  in LBCO.\cite{fujita04}  
Cuprate superconductors are expected to be in the vicinity of 
the $d$FSD instability, the so-called $d$-wave 
Pomeranchuk instability, leading to a giant response 
to a small external $xy$ anisotropy.

\begin{acknowledgments} 
The author is grateful to W. Metzner for collaboration on a 
related work and to R. Zeyher for a critical reading of the manuscript. 
\end{acknowledgments}


\bibliography{main.bib}

\newpage

\end{document}